\documentclass[%
reprint,
longbibliography,
nofootinbib,
amsmath,amssymb,
prd,
floatfix,
]{revtex4-2}

\usepackage{xcolor}
\usepackage{pifont}
\usepackage{graphicx}
\usepackage{dcolumn}
\usepackage{multirow}
\usepackage{bm}
\usepackage{hyperref}
\hypersetup{
  colorlinks   = true, 
  urlcolor     = blue, 
  linkcolor    = blue, 
  citecolor   = blue 
}

\AtBeginDocument{%
    \newwrite\bibnotes
    \def\bibnotesext{Notes.bib}
    \immediate\openout\bibnotes=\jobname\bibnotesext
    \immediate\write\bibnotes{@CONTROL{REVTEX42Control}}
    \immediate\write\bibnotes{@CONTROL{%
    apsrev42Control,author="48",editor="1",pages="1",title="0",year="1"%
    }%
    }
     \if@filesw
     \immediate\write\@auxout{\string\citation{apsrev42Control}}%
    \fi
}%
\begin{document}

\preprint{APS/123-QED}

\title{Impact of weak lensing on bright standard siren analyses}%

\author{Charlie T. Mpetha$^{1}$}
\email{c.mpetha@ed.ac.uk}
\author{Giuseppe Congedo$^{1}$}%
\author{Andy Taylor$^{1}$}%
\author{Martin A. Hendry$^{2}$}
\affiliation{%
$^{1}$Institute for Astronomy, School of Physics and Astronomy, University of Edinburgh,
Royal Observatory, Blackford Hill, Edinburgh, EH9 3HJ, United Kingdom\\
 $^{2}$SUPA, School of Physics and Astronomy, University of Glasgow, Glasgow G12 8QQ, United Kingdom}%

\date{\today}

\begin{abstract}
Gravitational waves from binary mergers at cosmological distances will experience weak lensing by large scale structure. This causes a (de-)magnification, $\mu$, of the wave amplitude, and a completely degenerate modification to the inferred luminosity distance $d_L$. The customary method to address this is to increase the uncertainty on $d_L$ according to the dispersion of the magnification distribution at the source redshift, $\sigma_{\mu}(z)$. But this term is dependent on the cosmological parameters that are being constrained by gravitational wave ``standard sirens", such as the Hubble parameter $H_0$, and the matter density fraction $\Omega_{\rm m}$. The dispersion $\sigma_{\mu}(z)$ is also sensitive to the resolution of the simulation used for its calculation. Tension in the measured value of $H_0$ from independent datasets, and the present use of weak lensing fitting functions calibrated using outdated cosmological simulations, suggest $\sigma_{\mu}(z)$ could be underestimated. This motivates an investigation into the consequences of mischaracterising $\sigma_{\mu}(z)$. We consider two classes of standard siren, supermassive black hole binary and binary neutron star mergers. Underestimating $H_0$ and $\Omega_{\rm m}$ when calculating $\sigma_{\mu}(z)$ increases the probability of finding a residual lensing bias on these parameters greater than $1\sigma$ by $1.5-3$ times. Underestimating $\sigma_{\mu}(z)$ by using low resolution$/$small sky-area simulations can also significantly increase the probability of biased results, the probability of a $1\sigma\,(2\sigma)$ bias in $H_0$ and $\Omega_{\rm m}$ found from binary neutron star mergers is $54\%\,(19\%)$ in this case. For neutron star mergers, the mean bias on $H_0$ caused by magnification selection effects is $\Delta H_0 = -0.1\,$km$\,$s$^{-1}\,$Mpc$^{-1}$. The spread around this mean bias---determined by assumptions on $\sigma_{\mu}(z)$---is  $\Delta H_0 = \pm 0.25\,$km$\,$s$^{-1}\,$Mpc$^{-1}$, comparable to the forecasted uncertainty. These effects do not impact merging neutron stars' utility for addressing the $H_0$ tension, but left uncorrected they limit their use for precision cosmology. For supermassive black hole binaries, the spread of possible biases on $H_0$ is significant, $\Delta H_0=\pm5\,{\rm km}\,{\rm s}^{-1}\,{\rm Mpc}^{-1}$, but $\mathcal{O}(200)$ observations are needed to reduce the variance below the bias. To achieve accurate sub-percent level precision on cosmological parameters using standard sirens, first much improved knowledge on the form of the magnification distribution and its dependence on cosmology is needed.

\end{abstract}

\maketitle
\section{Introduction}

Waves propagating through the Universe are magnified by the gravitational potential of large scale structure. This \textit{weak lensing} effect on light, observed through its effect on galaxy shapes, has been intensely studied as a source of cosmological information \cite{WL_B+S,WL_sys,kids,HSC,DESYR3}. Similarly to light, gravitational wave weak lensing (GW-WL) modifies the observed wave amplitude $h$, to $h' = h \sqrt{\mu}$, where a prime denotes a lensed quantity \cite{GW-WL_1st,GW-WL_2nd,GW-WL_3rd,GW-WL_4th,HH}. The magnification $\mu$ depends on the matter along the line of sight where $\mu=1$ corresponds to no effect, $\mu<1$ a net underdensity of matter, and $\mu>1$ a net overdensity.

The magnification distribution over the sky, $p(\mu,z)$, broadens towards higher redshifts, and is peaked at values below one as there are more voids than clusters. The tail of $p(\mu,z)$ is weighted towards values greater than one since clustered matter can have a larger effect on the magnification than voids. Gravitational waves from low redshift merging binaries, such as those observed using the LIGO-Virgo-Kagra (LVK) network \cite{GW_1st,LIGO_FUTURE,KAGRA,LIGO_INDIA}, will not experience impactful weak lensing, though they may still be strongly lensed \cite{SL,SL-GW-H0}. The space-based GW detector LISA \cite{LISA} expects to observe between a few and a few tens of super massive black hole binary (SMBHB) mergers in the mid-2030's over its nominal 5-year survey time with a duty cycle of $80\%$ \cite{Klein,LISA_MBHB,Mangiagli}. The effect of weak lensing on these observations will be significant, even for individual sources due to small uncertainties on their measured distances. A future network of $3^{\rm rd}$ generation ground-based detectors (3G), such as the Einstein Telescope \cite{ET} and Cosmic Explorer \cite{CE} is also anticipated to begin operations in the mid-2030's. These detectors will observe binary neutron stars (BNSs) at $z\!\sim\!2$ \cite{THESEUS_BSS2}. Weak lensing will not be as important for individual BNSs due to their larger distance uncertainties compared to those of SMBHBs.  On the other hand, with the 3G network expected to detect many more BNSs than the numbers of SMBHBs observed by LISA, the effect of instrumental scatter on the cosmological parameters will be reduced.  The bias due to lensing is likely to be even more significant.

Past works have studied how successfully a GW can be ``delensed"  using external data to estimate $\mu$ for each observed source \cite{Hilbert,Shapiro,Wu}. Ref.\;\cite{Wu} found that sufficient delensing using galaxy surveys for all SMBHB mergers observed with LISA will be extremely challenging. For both SMBHBs and BNSs, we can not expect to remove or even meaningfully reduce the weak lensing effect, unless further advances are made in delensing methods. An alternative delensing prospect which uses GW data alone is through measuring wave-optics features in a single waveform \cite{WOF}. However the expected probability for observing these features is low \cite{WOF2}, and this approach is not able to directly probe the range of lens masses which contribute most to weak lensing magnification.

Measuring the distance and redshift of a binary merger is the core of using these sources for cosmology, as ``standard sirens" \cite{Schutz}. Bright standard sirens (BSS) are the ideal case where the GW is observed alongside an electromagnetic counterpart. Another method for using standard sirens for cosmology utilises the weak lensing of a large number of GWs as information instead of noise, constraining structure as well as geometry \cite{ULTRA,GW_Cl_z_int,GCAT,Mukherjee_2020,Mukherjee_2020_B,Mpetha,balaudo,GWXEM}. But this is only viable with sufficient source numbers, likely not until at least the 2040s. In the meantime, many works have demonstrated how weak lensing can bias the distance redshift relation for the case of supernovae standard candles and GW standard sirens \cite{stoch_bias_0,stoch_bias_1,KP,SN_bias,SN_lensing,SS_lensing_PDF,Fleury_bias,bias_scatter,clumpy_lensing,stoch_bias_2,stoch_bias_3,stoch_bias_4,lensbias}. 

The luminosity distance $d_L$ is an observable quantity with gravitational waves from binary mergers, without needing any external calibration from the cosmic distance ladder. At redshift $z$,
\begin{equation}
    d_L(z) = (1+z)\int_0^z \frac{c \, dz'}{H(z')} \, ,\label{eq:dL}
\end{equation}
where the Hubble parameter is
\begin{align}
    \left(\frac{H(z)}{H_0}\right)^{2} &= \Omega_{\rm m}(1+z)^{3} + \Omega_{K}(1+z)^{2} \nonumber\\ &\quad + \Omega_{\rm DE}(1+z)^{3(1+w_0 +w_a)} e^{\frac{-3 w_a z}{1+z}} \, .
   \label{eq:Hz}
\end{align}

Here $H_0=100\,h\,{\rm km}\,{\rm s}^{-1}\,{\rm Mpc}^{-1}$ is the present expansion rate, $\Omega_{\rm m}$, $\Omega_{K}$ and $\Omega_{\rm DE}$ are the matter, curvature and dark energy density fractions respectively, and $w_0$ and $w_a$ are equation of state parameters for dark energy \cite{CPL1,CPL2}.

The luminosity distance is inversely proportional to the observed wave amplitude. Therefore the observed, lensed distance is
\begin{equation}
    d_L' = \frac{d_L}{\sqrt{\mu}} \, .
\end{equation}
The total distance uncertainty is
\begin{equation}
    \sigma_{d_L}^{2} = \sigma_{\rm GW}^{2} + \sigma_{\rm WL}^{2} + \left(\frac{\partial d_L}{\partial z}\right)^{2}\left(\sigma_{z}^{2} + \sigma_{v_{\rm pec}}^{2}\right) \, , \label{eq:sigtot}
\end{equation}
where $\partial d_L/\partial z$ converts redshift to distance uncertainties. $\sigma_{\rm GW}$ and $\sigma_{v_{\rm pec}}$ are the uncertainty from the detector and source peculiar velocity respectively. The weak lensing term is
\begin{equation}
    \frac{\sigma_{\rm WL}(z)}{d_L(z)} = \frac{1}{2}\frac{\sigma_{\mu}(z)}{\langle\mu(z)\rangle} = \frac{1}{2}\sigma_{\mu}(z)\, , \label{eq:sigwl}
\end{equation}
$\sigma_\mu(z)$ is the uncertainty in the magnification of the gravitational wave. It has been assumed that the mean of the magnification distribution $\langle\mu\rangle=1$. For real observations this is in general not the case due to detector selection effects as explored in Ref.\;\cite{lens_sel}, henceforth \hyperlink{cite.lens_sel}{CT21}.  The consequence is $\langle\mu\rangle\!\sim\!1.02$ at $z=2$.

Weak lensing could lead to a net over- or under-estimation of luminosity distances at different (true) redshifts, so that the inferred cosmological parameters in Eq.\;(\ref{eq:Hz}) are prone to biases \cite{lensbias}. A mitigation might be to include a weak lensing uncertainty according to Eq.\;(\ref{eq:sigwl}), using an assumed functional form for  $\sigma_\mu(z)$ \cite{ETWL,LISA_sigWL,Speri,lens_sel,pmu_fit, dance}. The possible lensing bias is washed out by increasing distance errors. A caveat of this added weak lensing uncertainty is that the form of the magnification function is not well known, is dependent on the uncertain values of cosmological parameters, and is affected by poorly understood physics influencing the growth of structure on small scales. There is tension in the value of $H_0$ \cite{H0_tension} (recently exacerbated by JWST observations \cite{H0_JWST}), and present weak lensing fitting functions have been created with cosmological simulations that could be missing large and small-scale power important in estimating $\sigma_{\mu} (z)$ \cite{TK11} (henceforth \hyperlink{cite.TK11}{T11}). Both of these facts motivate an investigation into how under- or overestimating the weak lensing uncertainty can impact a cosmological analysis using GWs. 

We will evaluate the \textit{residual lensing bias}, the lensing bias remaining even after including a weak lensing uncertainty term. Ref.\;\cite{lensbias} found the residual lensing bias for one idealised source population, demonstrating for the first time that the lensing bias can be greater than the statistical uncertainty for BNS mergers. Our goals differ from that work in that we are broadly assessing the impact of \textit{mischaracterising} the weak lensing uncertainty term for realistic observations from future space and ground-based GW detectors.

We find that underestimating the weak lensing uncertainty by using incorrect values for cosmological parameters leads to a factor $1.5$ $(3)$ increase in the probability of obtaining results biased by more than $1\sigma$ for BNSs (SMBHBs). We also demonstrate the importance of the resolution of simulations used to determine $\sigma_{\mu}(z)$, pointing out that missing power due to simulations not including very small and large scale magnification power could mean up to a factor of $10$ greater probability of obtaining results biased by more than $1\sigma$. For constraints from BNSs, we find the bias on $H_0$ caused by lensing is too small to affect their application to the $H_0$ tension. But the bias is at a similar level to the variance and so limits their use for precision cosmology. For SMBHBs the lensing bias is significant with a range $\sim\!5\,{\rm km}\,{\rm s}^{-1}\,{\rm Mpc}^{-1}$, but will only become relevant with $\mathcal{O}(200)$ observations. If GWs from merging binaries are to be used to their full potential by the next generation of detectors, much improved knowledge of $\sigma_\mu(z)$ is needed.

Section \ref{sec:method} describes the method used to evaluate the residual lensing bias. In Section \ref{sec:cats} we introduce the binary merger catalogues used in our analysis.
Section \ref{sec:mag} details the generation of magnification distributions and weak lensing uncertainties, and Section \ref{sec:results} gives our results. We conclude in Section \ref{sec:concl}.

\section{Method \label{sec:method}}

Our goal is to evaluate the probability of finding cosmological constraints that are biased due to lensing, in particular the \textit{residual lensing bias} caused by underestimating the weak lensing uncertainty. The following steps outline how we evaluate this residual lensing bias.

\begin{enumerate}
    \item Set a true cosmology by choosing values for cosmological parameters in Eq.\;(\ref{eq:Hz}).
    \item Choose a \textit{population}: supermassive black hole binaries (SMBHBs) with a particular seed model observed with LISA \cite{Mangiagli} (see  Section \ref{sec:LISA}), or binary neutron stars (BNSs) observed with third generation ground based detectors (see Section \ref{sec:3G}). Steps $3-7$ are repeated over many catalogue \textit{realisations}: independent sets of observations drawn from the population model. 
    \item Choose a catalogue realisation. For each merger in the catalogue, randomly draw a magnification $\mu$ as described in Section\;\ref{sec:mag} from the redshift and cosmology dependent magnification distribution. Applying the magnification to the measured merger distance and its uncertainty $\sigma_{\rm GW}$ (a more magnified source has a higher SNR and smaller uncertainty, see Appendix \ref{app:sigdL}) gives a lensed source, to be compared to the unlensed case. The unlensed case has $\mu=1$ for all sources. Only sources with a SNR greater than a chosen threshold after applying the magnification go into the catalogue.
    \item For each merger generate an observed distance and redshift by drawing from a Gaussian distribution with a width equal to the instrumental uncertainty and mean equal to the true distance$/$redshift. Also add scatter to the redshift caused by peculiar velocities, motion of the binaries not due to Hubble flow. This is done for both the lensed and unlensed sources, with each having the same random draw to ensure we are only comparing the effects of lensing ($d_{L}'^{ {\rm obs}} = d_{L}^{\rm obs} / \sqrt{\mu}$).
    \item In both the lensed and unlensed cases include a weak lensing uncertainty $\sigma_{\rm WL}(z)$, as in Eq.\;(\ref{eq:sigtot}), to the distance. The assumptions going into generating $\sigma_{\rm WL}(z)$ are the focus of this investigation.
    \item Given the observed values of $d_L$ and $z$ for each merger in the catalogue, perform Monte Carlo Markov Chain sampling of $d_L-z$ in Eq.\;(\ref{eq:dL}) to reconstruct cosmological parameters in the lensed and unlensed cases. In this work we use the \textsc{emcee} package \cite{emcee}.
    \item The absolute bias is the difference in the best fit recovered parameter value in the lensed and unlensed case. The bias in units of uncertainty is found from the parameter differences method of the \textsc{tensiometer} package\footnote{\url{https://github.com/mraveri/tensiometer}} \cite{tension1,tension2}. This uses information from the full $h-\Omega_{\rm m}$ 2D posterior to quantify the size of the bias, and does not assume Gaussianity in the parameter space.
    \item Aggregate the biases for each catalogue realisation of a population. The resulting distribution of biases from all realisations characterises the residual lensing bias that can be expected for that population, given the true cosmology and assumed weak lensing uncertainty.
    \item Repeat steps $3-8$ a further $\sim\!300$ times, and combine the distributions from each repetition. There is possible variation in observations due to the random drawing of magnifications and random scatter caused by distance and redshift uncertainties. By doing this, we have ensured we sample over both statistical variation in the catalogues, and statistical variation in the observations.
\end{enumerate}

\section{Binary Merger catalogues \label{sec:cats}}

\begin{figure}
    \centering
    \includegraphics{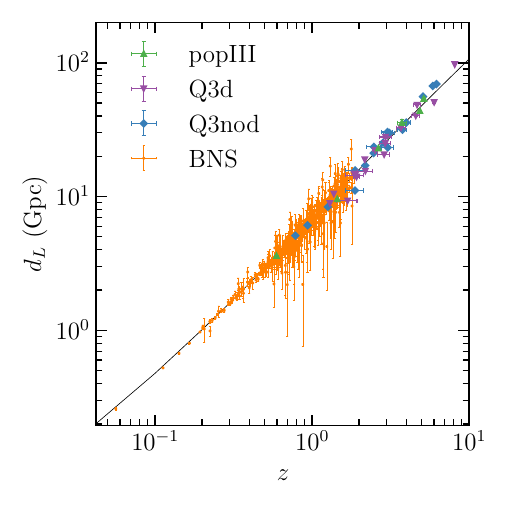}
    \caption{Representative catalogues for each binary merger population used in this work. Every gravitational wave has been observed alongside an electromagnetic counterpart. Sources have instrumental uncertainty and scatter in distance and redshift, and scatter caused by weak lensing and peculiar velocity. Weak lensing causes large scatter towards high redshift. The popIII light-seed SMBHB model typically predicts fewer, lower redshift mergers than the Q3 heavy-seed models.}
    \label{fig:dL-z}
\end{figure}

Our goal is to evaluate the residual lensing bias on cosmological parameters constrained with GWs. To this end, realistic catalogues of observations by the next generation of gravitational wave detectors are needed. For the following statistical tests $100/500$ catalogue realisations are used for BSS observed with LISA$/$3G. Section \ref{sec:LISA} details the subset of SMBHB catalogues from Ref.\;\cite{Mangiagli} used in this work, while the BNS catalogues we generated are described in Section \ref{sec:3G}. An example of one of the catalogues for each population, including instrumental distance$/$redshift uncertainties and magnification and instrumental scatter, can be seen in Fig.\;\ref{fig:dL-z}. The number of observations will depend on many factors; the performance of future GW and EM detectors, the fraction of the sky area covered by electromagnetic detectors for the multimessenger follow-up, the merger rate, and properties of the EM counterpart. For this reason we remain agnostic of rates and integration times, and instead use a broad range of source numbers in each catalogue. We choose not to consider dark standard sirens in this analysis, for example 3G observed binary black holes, or extreme-mass-ratio inspirals observed with LISA. The large redshift uncertainty of these sources will dominate over the weak lensing uncertainty for most cases. While Ref.\;\cite{dss_new} demonstrate that using several hundred 3G-observed ${\rm SNR}>300$ binary black holes could achieve comparable precision on $H_0$ to bright standard sirens, most of these will be too low redshift for important magnification effects.

All population catalogues have been generated assuming a Planck-like cosmology \cite{Planck}. In this work we evaluate the residual lensing bias in different true cosmologies, and while the assumed cosmology would impact the source distributions and uncertainties, we note that this effect is subdominant to details on the seed model and host galaxy evolution.

\subsection{LISA: Supermassive black hole binaries \label{sec:LISA}}

We use a subset of the catalogues presented in Refs\;\cite{Mangiagli,Mangiagli_forecasts}, refer there for more detail on the catalogue generation, and to Ref.\;\cite{Mangiagli_forecasts} for cosmological forecasts using these catalogues. As is common in the literature, uncertainty on the seed model for supermassive black holes motivates including three possibilities; black holes are formed from high-redshift high-mass metal poor ``popIII" stars \cite{popIII}, the ``Q3" heavy seed model where massive black holes are formed from the collapse of proto-galactic disks \cite {Q3_1,Q3_2} with a time delay between the formation time of a binary and its merger (Q3d), and the Q3 model with no delay time between formation and merger (Q3nod). Only mergers with an electromagnetic counterpart are considered. Other methods of estimating the redshift of a single source often rely on galaxy catalogues which will not be possible for high-redshift SMBHBs. In these catalogues there are three possibilities for the redshift uncertainty depending on the magnitude and redshift of the host galaxy observed with the ELT\footnote{\url{https://elt.eso.org/}}. If the host galaxy magnitude $m_{\rm gal,ELT} < 27.2$  and $0.5 < z_{\rm gal} \leq 5$ then $\sigma_z = 5\times10^{-4}$. If $27.2 < m_{\rm gal,ELT} < 31.3$ and  $0.5 < z_{\rm gal} \leq 5$  ($z_{\rm gal} > 5$) then $\sigma_z = 0.1$ ($\sigma_z = 0.25$). See Ref.\;\cite{Mangiagli} for more detail. We also include a peculiar velocity drawn from a Gaussian centred at zero with a width of $500\,$km$\,$s$^{-1}$. Sources (from all seed models) range from $z=0.8$ to $z=9.2$, with distance uncertainties from $0.02\%$ to $30\%$ having a median value of $0.8\%/0.4\%$ for the popIII$/$Q3 seed models. The average number of sources (median redshift) is 7 (2.4), 15 (3.2) and 20 (2.9) for popIII, Q3d and Q3nod respectively.

\subsection{3G: Binary neutron stars\label{sec:3G}}

For $3^{\rm rd}$ generation ground-based detectors (3G), the most likely BSS are binary neutron star (BNS) mergers. It is not expected that many black hole-neutron star mergers will be observed alongside an EM counterpart \cite{BHNS_DSS}, so this population is not included. BNSs are generated using the assumptions in Table 1 of Ref.\;\cite{gwfast}, which involve uniform sampling over sky location and inclination, uniform sampling of spins aligned with the orbital angular momentum in the range $[-0.05,0.05]$, and uniform sampling of mass in the range $[1,2.5]\,M_{\odot}$. The adopted redshift distribution of BSS is described in Ref.\;\cite{Mpetha}. The Fisher-matrix code \textsc{GWFAST} \cite{gwfast0} is used to find measurement uncertainties of each source in the catalogue. We assume a network of Einstein Telescope and two Cosmic Explorers (ET+2CE) for the main analysis, considering other configurations in Appendix \ref{app:dets}. 

$\mathcal{O}(10^5)$ BNSs will be observed every year by a 3G detector network \cite{listen,gwfast}. Predictions on the corresponding number of BSS vary significantly, from 15 yr$^{-1}$ to 3500 yr$^{-1}$ \cite{THESEUS_BSS,THESEUS_BSS2,SS_MG,MM_3G,lure,BNS_N,BSS_Aus} depending on assumptions on the GW and EM detectors, merger rates and counterparts. To reflect this our 500 generated catalogues have source numbers sampled uniformly between $100$ and $1000$. The minimum is chosen to be $100$ as this could be realised even with the lowest rate estimates after 10 years of runtime.

We impose a GW detector network SNR threshold of $\rho_{\rm lim}=12$ \cite{ET_SNR_12}. Due to the requirement of an EM counterpart, these observations come with selection effects beyond the requirements for a detection of the gravitational wave alone.  We add a cut on sky localisation of $20\,$deg$^{2}$ \cite{loc20,loc20_2}. We also limit the distance uncertainty to $\sigma_{d_L}/d_L<30\%$, around the upper limit of the $z - \sigma_{d_L}/d_L$ fitting function in Ref.\;\cite{ET_DE} for BSS observations with ET. As pointed out in Ref.\;\cite{HH}, knowledge of the sky location and inclination of a merger from its EM counterpart can improve constraints on GW parameters, quoting an order of magnitude improvement in $\sigma_{d_L}/d_L$. More recent studies use information on the inclination angle $\iota$ from EM observations to break the $d_L-\iota$ degeneracy and improve $\sigma_{d_L}/d_L$ by several factors \cite{viewing,afterglow,3G_dL_uncertainty}. This was performed for GW170817 using the radio counterpart,  improving the $H_0$ constraint by a factor of two \cite{GW170817_super}.  In the context of LISA, Ref.\;\cite{LISA_EM_local_unc} show how from sky localisation the wave amplitude and inclination uncertainties can both be improved by a factor of $\sim\!2$ when $\iota\nsim 90^o$. This factor increases to $\sim\!10$ if there is also some prior knowledge from EM observations that constrains the inclination. Therefore we make the simplifying assumption that, for all sources with $\iota\leq60^o$ and $\iota\geq120^o$ remaining after making the SNR and sky localisation cuts, the distance uncertainty is reduced by a factor of two using localisation information from the EM counterpart. We note that there could also be a constraint on the inclination angle for a fraction of observations, but do not explicitly include this effect. 

The final catalogues contain sources with redshifts from $0.018$ to $2$ and instrumental $d_L$ uncertainties ranging from $0.008\%$ to $30\%$, with a median value of $4\%$. Each merger has a peculiar velocity drawn from a Gaussian centred at zero with width $500\,{\rm km}\,{\rm s}^{-1}$. Finally, we assume spectroscopic redshift uncertainties of $\sigma_z = 5\times10^{-4}\,(1+z)$, based on electromagnetic follow up from present and forthcoming spectroscopic redshift surveys e.g DESI \cite{DESI}, ELT, and MegaMapper \cite{Megamapper}. The impact of this uncertainty is negligible.

\section{Magnification distributions \label{sec:mag}}

\begin{figure}
    \centering
    \includegraphics{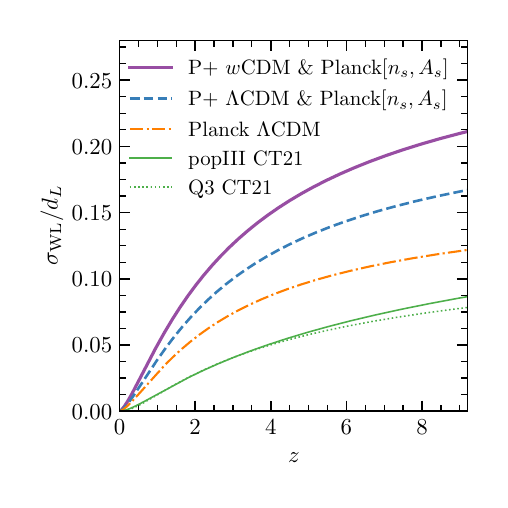}
    \caption{The weak lensing uncertainty fitting functions used in this work. Three have been created numerically in \textsc{camb} using different cosmologies: Planck 2018 \cite{Planck} and the Pantheon+ (P+) $\Lambda$CDM and $w$CDM maximum posterior values \cite{SNH0}. Supernovae only constrain geometry, so Planck 2018 values for $n_s$ and $A_s$ are used. Also shown are results from \protect\hyperlink{cite.lens_sel}{CT21} using numerical simulations in \protect\hyperlink{cite.TK11}{T11}, specifically for popIII\,/\,Q3 MBHBs observed with LISA, including selection effects.}
    \label{fig:sigmu}
\end{figure}

The following section details the redshift and cosmology dependent magnification distribution, and how it is used to generate a weak lensing fitting function. We will show how the width of the magnification distribution depends strongly on the assumed cosmological parameters and modelling assumptions. This lack of knowledge of the true form motivates an investigation into the effect on a cosmological analysis of mischaracterising the weak lensing uncertainty. The various weak lensing fitting functions used in this work are summarised in Fig.\;\ref{fig:sigmu}. 

\subsection{Background}
The magnification is given by
\begin{equation}
    \mu = \frac{1}{(1-\kappa)^2 - \gamma^2} \, .
\end{equation}
The convergence $\kappa$ is the integrated surface mass density along the line of sight, and $\gamma$ is the shear field. We are interested in the variance of the magnification at a particular redshift $\sigma_{\mu}^{2}(z)$, as this provides $\sigma_{\rm WL}(z)$ in Eq.\;(\ref{eq:sigtot}). For the rest of this work, unless otherwise stated, we will drop the small shear corrections which have little impact on gravitational waves \cite{GW_phase}. In that case, at a given redshift, the variance in the convergence field is directly related to the variance in the magnification \cite{dance}, 
\begin{equation}
    \sigma^2_{\mu}(z\,|\,\bm{\theta}) \approx 4 \sigma^2_{\kappa}(z\,|\,\bm{\theta}) \, .
\end{equation}
The convergence field is cosmology dependent. The cosmological parameters influencing the convergence field are $\bm{\theta}=\{h, \Omega_{\rm m}, \Omega_{\rm b}, \Sigma m_{\nu}, n_s, A_s, \Omega_{K}, w_0,w_a\}$. $\Omega_{\rm b}$ is the baryon density fraction, $\Sigma m_{\nu}$ the sum of neutrino masses, $A_s$ is the amplitude of scalar perturbations and $n_s$ the spectral tilt. Now, using Eq.\;(\ref{eq:sigwl}), the variance in the convergence field can be related to the weak lensing uncertainty we assume for a gravitational wave source,
\begin{equation}
    \frac{\sigma_{\rm WL}(z\,|\,\bm{\theta})}{d_L(z\,|\,\bm{\theta})} \approx \sigma_{\kappa}(z\,|\,\bm{\theta}) \, . \label{eq:WL_mu}
\end{equation}

Knowing the dispersion, a magnification probability distribution, $p(\mu,z\,|\,\bm{\theta})$, can be constructed. Poorly understood small-scale physics significantly impacts the high-magnification tail of $p(\mu,z\,|\,\bm{\theta})$. To avoid introducing modelling uncertainty, we restrict our analysis to the weak lensing regime, with $\mu \leq 1.75$\footnote{Imposing this limit causes $\langle\mu\rangle < 1$, but this effect is small ($0.01\%$ at $z=2$ and $0.5\%$ at $z=8$) and we confirmed it does not impact results for the size of the lensing bias.}. Therefore a simple shifted lognormal model as presented in Ref.\;\cite{dance} is sufficient, but for comparison we also use $p(\mu,z)$ from the $N_{\rm side}=4096$ and $N_{\rm side}=16384$ simulations described in Ref.\;\cite{TK}\footnote{\url{http://cosmo.phys.hirosaki-u.ac.jp/takahasi/allsky_raytracing/}} (henceforth \hyperlink{cite.TK}{T17}).

To include the impact of selection effects, we must modify the magnification probability distribution. Following Refs\;\cite{cusin2020,lens_sel},
\begin{equation}
    \mathcal{P}(\mu,z\,|\,\bm{\theta}) = \mathcal{C} \, p(\mu,z\,|\,\bm{\theta}) \frac{d\mathcal{N}(\mu,z)}{dz} \, ,
\end{equation}
where
\begin{equation}
    \frac{d\mathcal{N}(\mu,z)}{dz} = \int_{\rho_{\rm lim}/\sqrt{\mu}}^{\infty} N(z,\rho) \, d\rho \, .
\end{equation}
The SNR is given by $\rho$, $N(z,\rho)$ is the number density of sources as a function of redshift and SNR, $\rho_{\rm lim}$ is the SNR threshold and $\mathcal{C}$ is a normalisation constant. This modified magnification distribution is drawn from randomly to find the magnification of sources. Fig.\;16 of Ref.\;\cite{Mangiagli} shows that BSS SMBHBs have SNRs far above any possible SNR threshold, hence have no selection effects. This is not the case for BSS BNSs.

The dispersion of the magnification distribution as a function of redshift can be found using observations, simulations, or numerical calculation using a halo model. These possibilities are explored in the following sections.

\subsection{Numerical calculation \label{sec:numer}}

Ref.\;\cite{dance} provide a tool to generate the dispersion in a redshift and cosmology dependent convergence field, $\sigma_{\kappa}(z\,|\,\bm{\theta})$, for varying cosmological parameters using the Boltzmann code \textsc{camb}\footnote{\url{https://camb.info/}}. We made minor modifications to their notebook; using the HMcode-2020 nonlinear model \cite{hmcode}, setting $k_{\rm max} = 100\,h\,\rm{Mpc}^{-1}$ as this is where HMcode switches from calling the linear power spectrum to using a linear extrapolation, and including neutrinos and dark energy to the fitting. The main uncertainty in this approach is the modelling of non-linear scales. HMcode is known to not be reliable for $z>2$ \cite{hmcode}, and calculating the convergence power spectrum requires a power-law extrapolation of the Weyl potential at high $k-$modes which is not well motivated. This calculation involves integrating the convergence power spectrum over the multipole $\ell$ to find the total variance. This integral convergences for $\ell_{\rm max}=10^7$ \cite{dance}, and while we can not expect to accurately predict the power spectrum at such high multipoles, omitting small scales could lead to underpredicting the variance. Thus, knowledge of much smaller scales than presently available is needed for accurate modelling of $\sigma_{\kappa}(z\,|\,\bm{\theta})$.  

Using this modified notebook, $\sigma_{\kappa}(z\,|\,\bm{\theta})$ is found for different choices of cosmological parameters: Planck 2018 base $\Lambda$CDM constraints from TT,TE,EE+lowE+lensing+BAO \cite{Planck} (henceforth labelled `Planck'), and the best fit parameters from the Pantheon+ analysis assuming either $\Lambda$CDM or $w$CDM \cite{SNH0}. These choices highlight the current tension in constraints of $H_0$. Supernovae only constrain geometry so $n_s$ and $A_s$ are fixed to their Planck values when generating these fitting functions. The weak lensing uncertainty for various cosmologies is plotted in Fig.\;\ref{fig:sigmu}. The value of $h$, $\Omega_{\rm m}$, $w_0$ and $w_a$ used for each cosmology is given in Table\;\ref{tab:cosmo}.

\begin{table}
\caption{\label{tab:cosmo}
We will evaluate the residual lensing bias when one of these three cosmologies is used to generate a weak lensing fitting function, while the truth is given by another.}
\begin{ruledtabular}
\begin{tabular}{lllll} 
    Cosmology& $h$& $\Omega_{\rm m}$ & $w_0$&$w_a$ \\ 
    \colrule
     Planck& $0.677$& $0.311$& $-1$&$0$\\ 
     Pantheon+ $\Lambda$CDM& $0.736$& $0.344$& $-1$&$0$\\ 
     Pantheon+ $w$CDM& $0.733$& $0.403$& $-0.93$&$-0.1$\\ \end{tabular}
\end{ruledtabular}
\end{table}

\subsection{Simulations \label{sec:sim}}

The present industry standard weak lensing fitting function \cite{Hirata_sigWL_orig,LISA_sigWL,ETWL,Speri,lens_sel,pmu_fit} was created in 2010 \cite{Hirata_sigWL_orig} using the magnification distribution of the 1998 work Ref.\;\cite{HW}. It has since been modified in \hyperlink{cite.lens_sel}{CT21} using the ray-tracing simulations presented in \hyperlink{cite.TK11}{T11}. The use of these simulations gives cause for concern; they have limited resolution over a very small patch of the sky and therefore could under-predict the width of the magnification distribution, they are dark matter only simulations neglecting baryonic physics, and they were created using outdated values of cosmological parameters ($h=0.705$, $\Omega_{\rm m}=0.274$). Ref.\;\cite{lensing-pdfs-baryons} demonstrate how the inclusion of baryons in hydrodynamic simulations modifies the high magnification tail ($\mu > 3$), past the weak lensing regime. This suggests that dark matter only simulations are sufficient for evaluating the effects of weak lensing. But more recent works using the IllustrisTNG \cite{ITNG} and MilleniumTNG \cite{MTNG} simulation suites suggest that baryons, and in particular feedback effects, have a larger impact, causing a narrower $p(\mu,z)$ as matter is redistributed from high to low density regions. Furthermore, the magnification distribution is highly dependent on cosmological parameters. Lower values of $h$ and $\Omega_{\rm m}$ predict a smaller $\sigma_{\kappa}(z\,|\,\bm{\theta})$. More recent constraints on $\Omega_{\rm m}$ favour larger values \cite{Planck,SNH0} than used in \hyperlink{cite.TK11}{T11}, and there is the possibility $h$ favours the near-Universe measurement \cite{H0licow,SNH0}. Therefore current weak lensing fitting functions in the literature could be underestimating the dispersion of $p(\mu,z\,|\,\bm{\theta})$.

The function calibrated using \hyperlink{cite.TK11}{T11}, as presented in \hyperlink{cite.lens_sel}{CT21}, where the authors consider lensing selection effects for the popIII and Q3 SMBHB seed models separately, is plotted as a green solid and green dotted line in Fig.\;\ref{fig:sigmu}. The fitting function, whose form was originally found in Ref.\;\cite{Hirata_sigWL_orig}, is given by
\begin{equation}
        \frac{\sigma_{\rm WL}}{d_L} = \frac{C}{2} \left(\frac{(1 - (1 + z)^{-\beta})}{\beta}\right) ^ \alpha \, , \label{eq:sigWL_fit}
\end{equation}
where $C=0.061$, $\beta=0.264$, $\alpha=1.89$ for the popIII seed model and $C=0.096$, $\beta=0.62$, $\alpha=2.36$ for the Q3 model. Note the factor of $1/2$, which is missing in Ref.\;\cite{LISA_sigWL}. This is because the original expression is for the uncertainty in $\sigma({\rm ln}\,d_L^2) = 2\sigma(d_L)/d_L$\footnote{Incidentally, including this factor of $1/2$ would improve results in numerous forecasting papers, e.g. Refs\;\cite{LISA_sigWL,GCAT,Speri,SS_MG,SS_MG1,Mpetha}.} \cite{lens_sel,dss_new}. 

The difference between these curves and those generated using \textsc{camb} is significant. This is due to the simulation's resolution. As previously mentioned, for the numerical calculation there is an integral over the convergence power spectrum from $\ell_{\rm min}=1$ to $\ell_{\rm max}=10^7$. The ray-tracing simulations of \hyperlink{cite.TK11}{T11} use a $1\times1\,$deg$^2$ box with a grid resolution of $3\,$kpc$/h$. The small box will fail to include super-clusters or huge voids, both of which will broaden the magnification distribution. This sets a high $\ell_{\rm min}$, while $\ell_{\rm max}$ increases with redshift but is smaller than $10^7$. Hence the variance found from these simulations will be smaller. Setting $\ell_{\rm min}$ and $\ell_{\rm max}$ in the numerical calculation based on the simulation parameters leads to comparable predictions for $\sigma_{\rm WL}(z)$ to  Ref.\;\cite{Hirata_sigWL_orig} and \hyperlink{cite.lens_sel}{CT21}.

\begin{figure}
    \centering
    \includegraphics{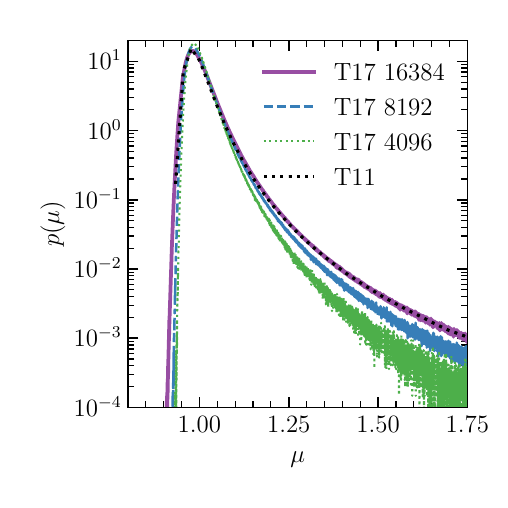}
    \caption{The magnification probability distribution at $z=1.0334$ found from the full-sky ray-tracing simulations of \protect\hyperlink{cite.TK}{T17}. Increasing the $N_{\rm side}$ of the simulation from $4096$ to $16384$ leads to a broader distribution, and therefore a larger assumed weak lensing uncertainty. Also overlaid is the $z=1$ pdf from the \protect\hyperlink{cite.TK11}{T11} $1\times1\,$deg$^2$ ray-tracing simulations.}
    \label{fig:TK_Nside}
\end{figure}

\begin{figure}
    \centering
    \includegraphics[width=\linewidth]{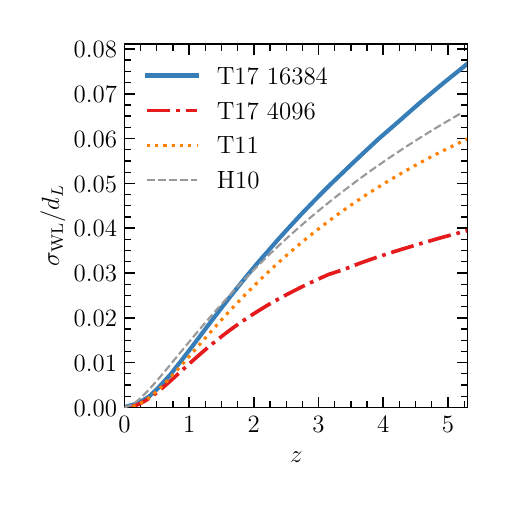}
    \caption{A comparison of weak lensing uncertainties from full-sky ray tracing simulations with different resolutions (\protect\hyperlink{cite.TK}{T17}), high resolution ray-tracing simulations over a $1\times1\,$deg$^2$ patch of sky (\protect\hyperlink{cite.TK11}{T11}), and from the analytic pdf in Ref.\;\cite{HW} used in Ref.\;\cite{Hirata_sigWL_orig} (H10). No selection effects have been included.}
    \label{fig:sigwl_comparison}
\end{figure}

One could also consider using the ray-tracing simulations of \hyperlink{cite.TK}{T17} to find the dispersion in the magnification distribution over finer redshift slices.  They are full-sky, providing access to lower multipoles, but have significantly lower resolution in the high $\ell$ regime, and are available up to a maximum redshift of $5.3$, compared to $z_{\rm max}=20$ in \hyperlink{cite.TK11}{T11} (though there are only 7 redshift bins from $z=1$ to $z=20$). They use a similar set of cosmological parameters to \hyperlink{cite.TK11}{T11}, with $h=0.7$, $\Omega_{\rm m}=0.279$. To demonstrate the importance of simulation resolution, the magnification distribution from simulations of \hyperlink{cite.TK}{T17} for the same snapshot at $z=1.0334$, but with a different $N_{\rm side}$, can be seen in Fig.\;\ref{fig:TK_Nside}. Both the convergence and shear fields are used to find $p(\mu)$ from these simulations. There is a marked difference in the width of $p(\mu)$, highlighting the importance of resolution (or equivalently, $\ell_{\rm max}$) in estimating the weak lensing uncertainty. Also shown is $p(\mu)$ from \hyperlink{cite.TK11}{T11} at $z=1$, which shows close similarity to the $N_{\rm side}=16384$ case. This suggests that even finer resolution full-sky simulations could give a broader distribution than found in \hyperlink{cite.TK11}{T11}. Corresponding weak lensing uncertainties found using the method of Ref.\;\cite{Hirata_sigWL_orig} can be seen in Fig.\;\ref{fig:sigwl_comparison}. The weak lensing uncertainty found from the full-sky \hyperlink{cite.TK}{T17} $N_{\rm side}=16384$ simulations follows the fitting function of Eq.\;(\ref{eq:sigWL_fit}) with $C=0.076$,  $\beta=0.344$ and $\alpha=2.25$ with no selection effects, and $C=0.092$,  $\beta=0.164$, $\alpha=2.09$ considering selection effects from a ET+2CE network. The finer redshift slices and full-sky nature make this fitting function more appropriate for the redshift regime where it is applicable, $z \leq 5.3$. We suggest this fitting function should be used for the lensing uncertainty on BNS mergers. However higher resolution full-sky simulations will likely lead to a wider predicted magnification distribution.

Weak lensing data products from other cosmological simulations exist \cite{abacus,Harnois_WL,Buzzard,cosmodc2,cosmogrid,ELUCID} and it would be interesting to see how choice of simulation affects the derived weak lensing uncertainty. To our knowledge, the only other example of a weak lensing uncertainty predicted from simulations is shown in Ref.\;\cite{ETWL}. The authors use the ELUCID simulations \cite{ELUCID} to estimate $\sigma_{\rm WL}(z)$ for observations of binary black holes at $z \leq 1$. For $z > 1$ they also use results from \hyperlink{cite.TK11}{T11}, hence find general agreement with the curves in \hyperlink{cite.lens_sel}{CT21}.

\subsection{Observations}

Ideally, $p(\mu, z)$ would be inferred from observations. Through the weak lensing of galaxies we measure reduced shear $g=\gamma/(1-\kappa)$, which is either approximated as shear, or $\gamma$ is recovered through an iterative scheme \cite{KS+}. The convergence is then reconstructed from the shear. This is a very noisy process, and measuring enough galaxies for accurate reconstruction requires deep observations. Measuring $p(\mu,z)$ directly over the full sky in fine redshift slices is not achievable in the near future. The best we can do is calibrate simulations to lensing statistics from forthcoming large scale galaxy surveys,  such as the Vera C. Rubin Observatory \cite{LSST} or Euclid \cite{EUCLID}.
 
\section{Results \label{sec:results}}

\subsection{Evaluating the residual lensing bias \label{sec:eval}}

For each combination of true and assumed cosmology, the steps in Section \ref{sec:method} are performed to gain a large distribution of possible lensing biases for each population. Using this distribution, we can characterise the impact of weak lensing through the mean bias, and the dispersion of biases. Positively magnified sources are more likely to be observed, this is a weak lensing selection effect. When there are no selection effects $\langle\mu\rangle=1$ and the mean bias$\,\sim0$. But with low SNR sources that can be shifted inside the SNR limit through a weak lensing magnification, $\langle\mu\rangle\neq1$, and there is a bias in the recovered cosmological parameters. This depends primarily on the source redshift distribution and the magnification distribution. For a chosen population, the mean residual lensing bias found by averaging over all statistical samples of a single catalogue realisation shows little variation from realisation to realisation. Even though the realisations have a range of source numbers, this will only impact the parameter uncertainties and not the mean lensing bias. For the BNS catalogues considered in this work, we report the mean absolute bias on $h$ and $\Omega_{\rm m}$  due to selection effects in Table\;\ref{tab:abs_bias}. Results are shown for the three cosmologies used to generate the numerical estimate of $\sigma_{\rm WL}(z)$, and when magnifications are drawn directly from the $N_{\rm side}=4096$ and $N_{\rm side}=16384$ simulations of \hyperlink{cite.TK}{T17}. In the $N_{\rm side}=4096$ case the mean bias is smaller as the limited resolution leads to a smaller $\ell_{\rm max}$ than in the $N_{\rm side}=16384$ and numerical cases. BSS SMBHBs need a high SNR to accurately localise and associate with an EM counterpart so there are no magnification selection effects on these sources, and their mean bias fluctuates around zero across each catalogue.

\begin{table}
\caption{\label{tab:abs_bias}
The magnification selection effect on gravitational waves causes a bias $\overline{b}$ on recovered cosmological parameters. This depends on the source redshift distribution: results for the BNS catalogues used in this work are given here. The bias has been averaged over $300$ statistical realisations of $500$ catalogues. The bias also depends on the width of magnification distribution, which is affected by cosmological parameters and simulation resolution.}
\begin{ruledtabular}
\begin{tabular}{lll} 
    $p(\mu)$ Cosmology& $\overline{b}(h)$& $\overline{b}(\Omega_{\rm m})$\\ 
    \colrule
     Planck Log Norm & $-0.0011$& $0.0074$\\ 
     Pantheon+ $\Lambda$CDM Log Norm& $-0.0015$& $0.0098$\\ 
     Pantheon+ $w$CDM Log Norm & $-0.0020$& $0.0153$\\
 \hyperlink{cite.TK}{T17} ($N_{\rm side}=4096$)& $-0.0007$&$0.0044$\\ 
 \hyperlink{cite.TK}{T17} ($N_{\rm side}=16384$)& $-0.0013$ & $0.0083$ \\ \end{tabular}
\end{ruledtabular}
\end{table}

Next, we go on to consider the impact of the assumption on $\sigma_{\rm WL}(z)$. This will primarily affect the width of the bias distribution. From Table\;\ref{tab:abs_bias} it can already be seen how mischaracterising the true cosmology, or using low resolution simulations, when correcting for magnification effects could lead to a residual lensing bias. For the SMBHB catalogues, the high redshift sources lead to large magnifications and large absolute biases. But the small source numbers mean large parameter uncertainties as the bias is below the level of the variance. For BNSs the opposite is true: they have both smaller magnifications and uncertainties. The absolute bias is small, but its impact is larger. We quote results in two ways, first the probability of the bias on $h$ ($\Omega_{\rm m}$) exceeding a value of $\pm 0.0025$ ($\pm 0.015$) for BNSs and $\pm 0.02$ $(\pm 0.05$) for SMBHBs. This indicates the spread of biases across catalogues. Then the probability of there being a discordance in the parameter space greater than a certain number of standard deviations. This takes into account the parameter uncertainties to show the impact of the residual lensing bias on parameter constraints.

\subsubsection{No added weak lensing uncertainty}
First we will evaluate the impact of not including a weak lensing uncertainty term in Eq.\;(\ref{eq:sigtot}). If this term is not added, the probability of a lensing bias is given in Table \ref{tab:sigs_uncorrected}, where the true cosmology is Planck. It can be seen that the probability of obtaining highly significant biases from LISA sources is very large, due to the small uncertainties on the high redshift, high SNR SMBHBs and the lack of a weak lensing uncertainty term. Even for the ground-based observed BNSs the probability is high, demonstrating that weak lensing is an important consideration for these sources.

\begin{table}
\caption{\label{tab:sigs_uncorrected}
Mean probability of finding a lensing bias when a lensing uncertainty term on the distance is not included. Both the probability of the bias on $h$ and $\Omega_{\rm m}$ exceeding a set value, and the probability of the bias exceeding a number of standard deviations in the 2D $h-\Omega_{\rm m}$ posterior, are shown.  The true cosmology used is Planck 2018 \cite{Planck}. For BNSs $x_h=0.0025$, $x_{\Omega_{\rm m}} = 0.015$. For SMBHBs $x_h=0.02$, $x_{\Omega_{\rm m}} = 0.05$. }
\begin{ruledtabular}
\begin{tabular}{llll}
 Population  &  $P(b(h)\geq\pm x_h)$& $P(b(\Omega_{\rm m})\geq\pm x_{\Omega_{\rm m}})$&$P(b\geq1\sigma [3\sigma])$\\
     \colrule
     popIII  & $0.65$& $0.72$& $0.90$ $[0.69]$\\
     Q3  & $0.78$& $0.77$& $0.98$ $[0.79]$\\
     Q3nod  & $0.56$& $0.62$&  $0.98$ $[0.81]$\\
     BNS & 0.62& 0.55& $0.79$ $[0.44]$\end{tabular}
\end{ruledtabular}
\end{table}

This scenario highlights the necessity of including $\sigma_{\rm WL}(z)$. Any cosmological analysis using well measured binary mergers beyond $z\!\sim\!0.5$ that does not could be significantly biased.

\subsubsection{Correct uncertainty implemented \label{sec:correct_var}}
Now we test the situation where $\sigma_{\rm WL}(z)$ is generated using the same distribution that magnifications are drawn from. There is perfect knowledge of the true $p(\mu,z\,|\,\bm{\theta})$. This does not mean the magnification can be removed from the GW, for this an external measurement of $\mu$ is needed for each source. Magnifications are drawn from $p(\mu,z\,|\,\bm{\theta})$ generated using Planck parameters and the added uncertainty is the Planck curve in Fig.\;\ref{fig:sigmu}. Results are presented in Table \ref{tab:sigscorr}. 

In all populations, the probability of a large parameter shift remains high. The absolute bias due to lensing is primarily dependent on the shape of the magnification distribution, which is the same as it was in Table \ref{tab:sigs_uncorrected}.

In the case of SMBHBs, for a bias $b\geq1\sigma$ the probability is $\leq0.05$, and for $b\geq2\sigma$ the probability is $\leq0.01$. The lensing bias has been effectively removed by increasing uncertainties according to the dispersion of $p(\mu,z\,|\,\bm{\theta})$. 

For BNSs  we are more likely to observe sources with a positive magnification. There is an average negative$/$positive shift on $h/\Omega_{\rm m}$, also shown in Ref.\;\cite{lensbias}.  Due to this selection effect the probability of a residual bias $b\geq1\sigma\,(b\geq2\sigma)$ is $0.18\,(0.01)$. The lensing selection effect could in principle be corrected for using the method of \hyperlink{cite.lens_sel}{CT21}, outlined in their Eq.\;(10). Following this we confirmed that, if source numbers are large enough, binning them by redshift,  finding $\langle d_L^2(z)\rangle$ in each bin and multiplying by $\langle\mu(z)\rangle$ for that bin removed the lensing bias caused by selection effects. The key caveats of this method are needing sufficient source numbers to obtain a reliable $\langle d_L(z)\rangle$, and crucially, accurate knowledge of $p(\mu,z\,|\,\bm{\theta})$. As will be demonstrated in the next section, mischaracterising the magnification distribution could exacerbate the lensing bias instead of mitigating it.

\begin{table}
\caption{\label{tab:sigscorr}
Mean probability of finding a lensing bias. Both the probability of the bias on $h$ and $\Omega_{\rm m}$ exceeding a set value, and the probability of the bias exceeding a number of standard deviations in the 2D $h-\Omega_{\rm m}$ posterior, are shown. The weak lensing uncertainty used is based on perfect knowledge of the shape of the magnification distribution, created using Planck 2018 parameters. For BNSs $x_h=0.0025$, $x_{\Omega_{\rm m}} = 0.015$. For SMBHBs $x_h=0.02$, $x_{\Omega_{\rm m}} = 0.05$.}
\begin{ruledtabular}
\begin{tabular}{llll}
    Population  & $P(b(h)\geq\pm x_h)$&  $P(b(\Omega_{\rm m})\geq\pm x_{\Omega_{\rm m}})$&$P(b\geq1\sigma [2\sigma])$\\
         \colrule
         popIII & $0.45$ & $0.47$  &$0.03\,[0.00]$\\
         Q3 & $0.54$ &  $0.53$ &$0.03\,[0.00]$\\
         Q3nod & $0.32$ & $0.40$  &$0.05\,[0.01]$\\
         BNS  & $0.21$&  $0.14$&$0.18\,[0.01]$\end{tabular}
\end{ruledtabular}
\end{table}

\subsubsection{Imperfect correction}

\begin{figure}
    \centering
    \includegraphics{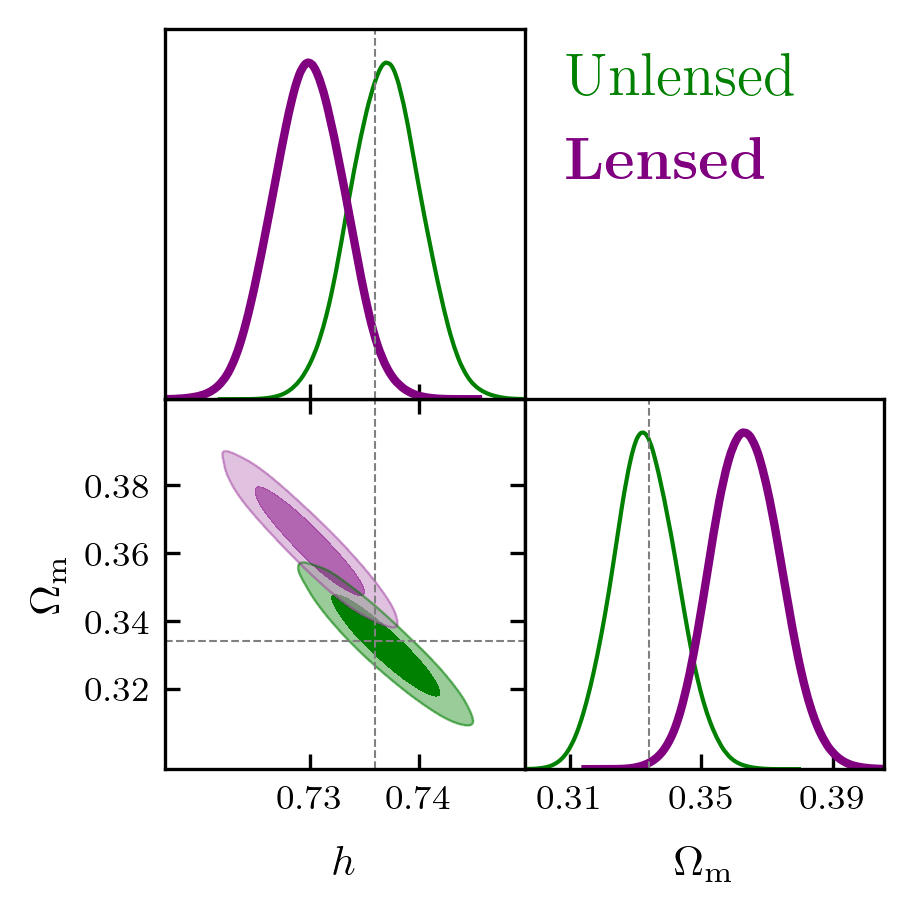}
    \caption{The Lensed and Unlensed contours are constraints from the same binary neutron star catalogue with $641$ sources. The \textit{true} cosmology is the the best fit Pantheon+ $\Lambda$CDM parameters. In both cases, the distance uncertainty is increased according to a weak lensing uncertainty fitting function generated using Planck 2018 parameters: this is the assumed cosmology. The green ``Unlensed" contour is the constraint on $h$ and $\Omega_{\rm m}$ when all sources experience magnification $\mu=1$. The purple ``Lensed" contour is when each source experiences a magnification drawn from a redshift dependent magnification distribution created using the \textit{true} cosmology. The parameters of this model give a broader magnification distribution than the weak lensing uncertainty fitting function predicts, hence the lensing bias is not completely mitigated; there is a total bias of $2.0\sigma$, or an absolute bias of $b(h)=-0.007$ and $b(\Omega_{\rm m})=0.031$.}
    \label{fig:lens_bias}
\end{figure}

In reality, we do not know the correct form of $p(\mu,z\,|\,\bm{\theta})$. First we will consider the case when the uncertainty is underestimated, and then when it is overestimated. If the weak lensing uncertainty is underestimated, then the lensing bias demonstrated in Table\;\ref{tab:sigs_uncorrected} will not be mitigated, as it has mostly been in Section\;\ref{sec:correct_var}. There are two main reasons the weak lensing uncertainty could be underestimated: 
\begin{itemize}
    \item[A.] Using too small values of $H_0$ or $\Omega_{\rm m}$ to predict $\sigma_{\rm WL}(z)$.
    \item[B.] Incorrect modelling of the shape of the magnification distribution.
\end{itemize} 

We investigate case A by assuming the true cosmology used to generate $p(\mu,z\,|\,\bm{\theta})$ is either the Pantheon+ $\Lambda$CDM or Pantheon+ $w$CDM best fit parameters, and the weak lensing uncertainty is given by the Planck curve in Fig.\;\ref{fig:sigmu}. Fig.\;\ref{fig:lens_bias} shows an example of the residual lensing bias from ET+2CE observations of BNSs with an electromagnetic counterpart. Probabilities of obtaining a lensing bias are presented in Table\;\ref{tab:sigsA}. A key result is the importance of accurate modelling of  $p(\mu,z\,|\,\bm{\theta})$ even for large numbers of BSS BNSs, where the lensing bias is greater than that caused by selection effects alone. We also tested the bias when BNSs are used to probe extensions to the base $\Lambda$CDM, including $k\Lambda$CDM (curvature) and $w_0$CDM (dark energy) in Appendix \ref{app:cosmo}. For LISA SMBHBs, despite the large probability of a significant parameter shift, the broad parameter uncertainties make the probability of a large discordance caused by lensing small.

\begin{table*}
\caption{\label{tab:sigsA}
Mean probability of finding a lensing bias. Both the probability of the bias on $h$ and $\Omega_{\rm m}$ exceeding a set value, and the probability of the bias exceeding a number of standard deviations in the 2D $h-\Omega_{\rm m}$ posterior, are shown. The weak lensing uncertainty is a fitting function based on the Planck 2018 cosmological parameters. The column headers give the true cosmology. For BNSs $x_h=0.0025$, $x_{\Omega_{\rm m}} = 0.015$. For SMBHBs $x_h=0.02$, $x_{\Omega_{\rm m}} = 0.05$.}
\begin{ruledtabular}
\begin{tabular}{lllllll}
    \multirow{2}{*}{Population}  &  \multicolumn{3}{c}{Pantheon+ $\Lambda$CDM}& \multicolumn{3}{c}{Pantheon+ $w$CDM}\\
         & $P(b(h)\geq\pm x_h)$&  $P(b(\Omega_{\rm m})\geq\pm x_{\Omega_{\rm m}})$&$P(b\geq1\sigma [2\sigma])$& $P(b(h)\geq\pm x_h)$&  $P(b(\Omega_{\rm m})\geq\pm x_{\Omega_{\rm m}})$&$P(b\geq1\sigma [2\sigma])$\\
         \colrule
         popIII & $0.55$&  $0.56$&$0.07\,[0.02]$& $0.62$&  $0.65$&$0.15\,[0.04]$\\
         Q3 & $0.66$&  $0.64$&$0.10\,[0.02]$& $0.72$&  $0.73$&$0.18\,[0.05]$\\
         Q3nod & $0.44$&  $0.53$&$0.13\,[0.02]$& $0.49$&  $0.64$&$0.21\,[0.05]$\\
         BNS & $0.32$&  $0.24$&$0.29\,[0.02]$& $0.50$&  $0.53$&$0.49\,[0.10]$\\
\end{tabular}
\end{ruledtabular}
\end{table*}

\begin{table}
\caption{\label{tab:sigsB}
Mean probability of finding a lensing bias. Both the probability of the bias on $h$ and $\Omega_{\rm m}$ exceeding a set value, and the probability of the bias exceeding a number of standard deviations in the 2D $h-\Omega_{\rm m}$ posterior, are shown. The true cosmology is given by the Planck parameters and the weak lensing uncertainty is given by \protect\hyperlink{cite.lens_sel}{CT21}. For BNSs $x_h=0.0025$, $x_{\Omega_{\rm m}} = 0.015$. For SMBHBs $x_h=0.02$, $x_{\Omega_{\rm m}} = 0.05$.}
\begin{ruledtabular}
\begin{tabular}{llll}
    \multirow{2}{*}{Population}  &  \multicolumn{3}{c}{Planck \& CT21}\\
         & $P(b(h)\geq\pm x_h)$&  $P(b(\Omega_{\rm m})\geq\pm x_{\Omega_{\rm m}})$&$P(b\geq1\sigma [2\sigma])$\\
         \colrule
         popIII & $0.53$ & $0.55$  &$0.32\,[0.09]$\\
         Q3 & $0.64$ &  $0.64$ &$0.38\,[0.11]$\\
         Q3nod & $0.37$ & $0.44$  &$0.46\,[0.15]$\\
         BNS  & $0.32$&  $0.25$&$0.54\,[0.19]$\end{tabular}
\end{ruledtabular}
\end{table}

To assess the impact of underestimating cosmological parameters, probabilities in Table\;\ref{tab:sigsA} are compared with those in Table\;\ref{tab:sigscorr}, which describe when the weak lensing uncertainty is correctly estimated. First consider if the true cosmology is Pantheon+ $\Lambda$CDM, and assumed is Planck. For all populations, the probability of the absolute bias on $h$ and $\Omega_{\rm m}$ of exceeding a certain value has increased. This is due to the wider magnification distribution leading to more significant magnification effects. For SMBHBs, the probability of obtaining results biased by more than $1\sigma$ is increased by a factor $\sim\!2.5-3$, but still remains low due to the large parameter uncertainties caused by few observations. For BNSs, the probability of results biased by more than $1\sigma$ has increased by a factor of $1.5$, and is close to $1/3$. If the true cosmology is in fact Pantheon+ $w$CDM, probabilities increase by an extra factor of $\sim\!2$.

To test case B, we compare the fitting functions of \hyperlink{cite.lens_sel}{CT21} with curves created using \textsc{camb}. $\sigma_{\rm WL}(z\,|\,\bm{\theta})$ generated using the numerical method of Section\;\ref{sec:numer} and the cosmological parameters in \hyperlink{cite.TK}{T11} and \hyperlink{cite.TK}{T17} are very similar to the Planck curve in Fig.\;\ref{fig:sigmu} (orange dash-dot line). This is because the increase in $\Omega_{\rm m}$ and decrease in $h$ between the parameters used in the simulations and Planck are competing effects and there is a partial cancellation of biases. The difference in $\sigma_{\rm WL}$ is $\lesssim 5\%$, so using Planck as the true cosmology to draw magnifications, and the fitting functions of \hyperlink{cite.lens_sel}{CT21} (green solid and dotted lines) as the lensing uncertainty acts as a model comparison. The probability of obtaining a residual lensing bias in this case is presented in Table\;\ref{tab:sigsB} for each population. Once again, as the true cosmology is the same as in Section\;\ref{sec:correct_var}, we can compare bias probabilities with Table\;\ref{tab:sigscorr}. In this case the uncertainty is too small to wash out the bias. For SMBHBs the probability of results biased by $\geq1\sigma$ increases by a factor of $\sim\!10$, and there is now a significant probability of results biased by $\geq2\sigma$. For BNSs, there is now greater than $50\%$ probability of finding results biased by more than $1\sigma$ and for $b\geq2\sigma$ the probability has jumped from $1\%$ to $19\%$. Due to the impact of selection effects on BNSs and their stronger constraining power, combining the SMBHB and BNS catalogues has a small impact on the BNS results.

If the assumption on $\sigma_{\rm WL}(z\,|\,\bm{\theta})$ in fact overestimates the dispersion in $p(\mu,z\,|\,\bm{\theta})$, then the uncertainty on the cosmological parameters is excessively diluted. When the true cosmology is Planck and the assumed cosmology is Pantheon+ $\Lambda$CDM ($w$CDM), the increased uncertainty at low redshift causes a degradation of parameter constraints of $10\%\,(15\%)/8\%\,(12\%)$ for SMBHBs$/$BNSs.

\subsection{Reducing the probability of a lensing bias}

\subsubsection{Setting requirements}

In Section \ref{sec:eval} we calculated the residual lensing bias in several test cases, demonstrating the bias can exceed the variance significantly if either cosmological parameters are underestimated, or too-low resolution simulations are underestimating the weak lensing uncertainty. We now set requirements on the required precision's to which the input cosmology should be known, and discuss prospects for using state-of-the-art simulations to prevent the residual lensing bias becoming a limiting factor for bright standard siren analyses.

The dependence of the residual lensing bias on $H_0$ and $\Omega_{\rm m}$ to the biased input parameters can be easily tested in our framework. For a fixed true cosmology, the average lensing bias is constant as it only depends on the magnification and source distributions. When we vary the input cosmology used to calculate the weak lensing uncertainty, we are just scaling the average uncertainty on recovered cosmological parameters (smaller assumed $\sigma_{\rm WL}(z\,|\,\bm{\theta})$ leads to smaller parameter uncertainties). And so the scaling of the average bias in units of $\sigma$ will simply depend on the scaling of $\sigma$, the parameter uncertainties.  The reason we have cast the change of bias in this way is to remain agnostic of the true source numbers$/$distribution and the true cosmology. 

\begin{figure*}
    \centering
    \includegraphics[width=\linewidth]{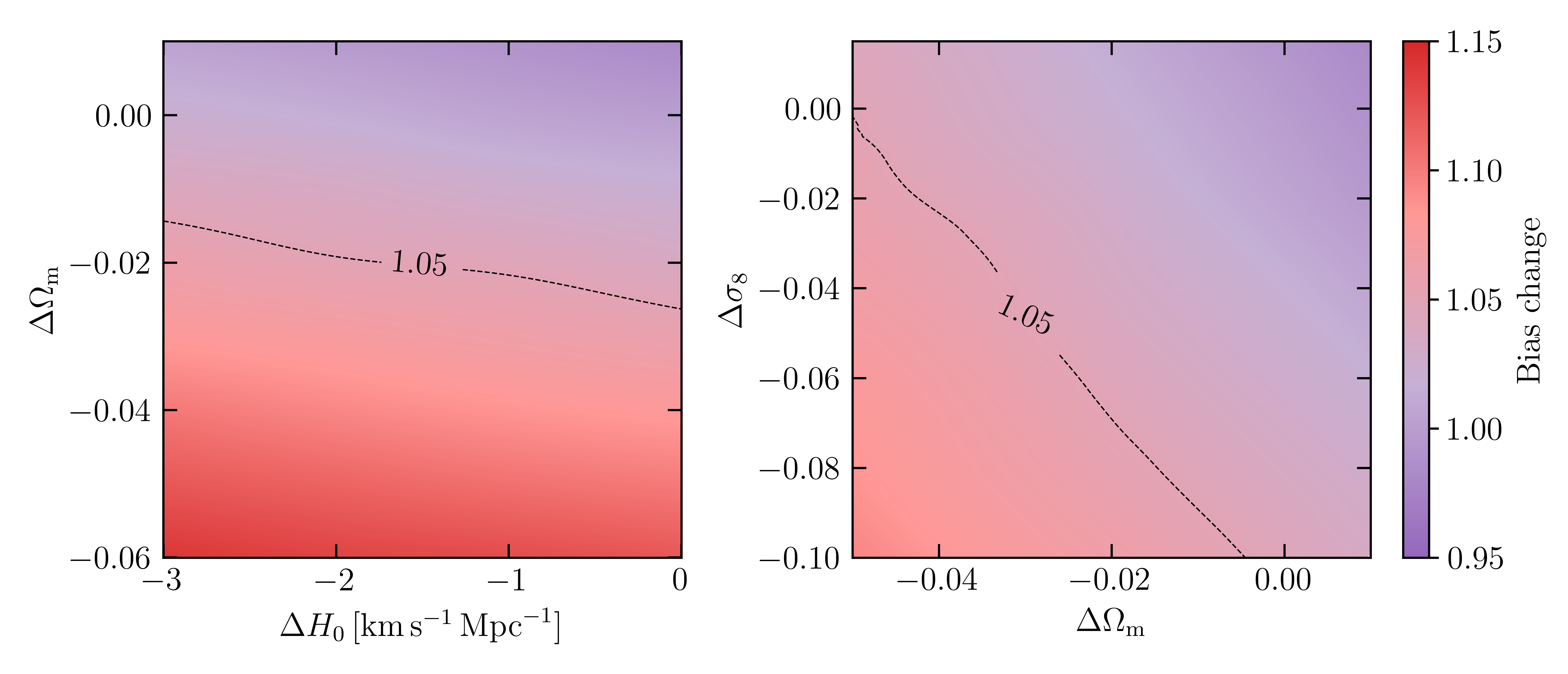}
    \caption{Heatmap displaying the lensing bias change due to incorrect cosmological parameters when generating the weak lensing uncertainty. The $x-$ and $y-$axes show the difference between the parameter used to generate the weak lensing uncertainty, and the true value of that parameter. The colour shows the resulting change in the average residual lensing bias, relative to the bias when the input cosmology is correct, $(0, 0)$ on these plots. Also overlaid is a contour showing the boundary where the change in the bias is at the $5\%$ level.}
    \label{fig:bias_dep}
\end{figure*}

In Fig.\;\ref{fig:bias_dep} we show the change in the lensing bias (the colour of the heatmap) as a function of the \textit{input bias} on cosmological parameters (the $x-$ and $y-$axes). The bias change is computed as a fractional change relative to the lensing bias when the input bias is zero, this is the $(0,0)$ point on these plots. When the input bias is zero, this can be considered a best-case scenario. If the input bias is negative, the lensing bias is increased and the bias change is greater than one. If the input bias is positive, then the bias change is less than one, but the uncertainties on the recovered $H_0$ and $\Omega_{\rm m}$ have been increased, weakening their constraints. We confirmed that the results in Fig.\;\ref{fig:bias_dep} are general for different source numbers and input cosmologies. The left plot shows that if the input $\Omega_{\rm m}$ is known to a precision of $0.02$ or better, the increase in bias is kept at or below the $5\%$ level. The $H_0$ offset is less impactful; weak lensing is less sensitive to the Hubble parameter \cite{Hall}, but $H_0$ should be known in combination with $\Omega_{\rm m}$ to $\sim\!2\,{\rm km}\,{\rm s}^{-1}\,{\rm Mpc}^{-1}$. In the right plot we also consider the amplitude of the linear matter power spectrum in spheres of radius $8\,{\rm Mpc}^{-1}\,h^{-1}$, $\sigma_8$. We find that, if $\Omega_{\rm m}$ is to be known to $0.02$ or better, then $\sigma_8$ should be known to a precision of $0.07$ to keep the increase in bias below the $5\%$ level.  The required precision's for $\sigma_8$ and $\Omega_{\rm m}$ are well within parameter uncertainties from e.g. Planck,  and are consistent with differences in these parameters observed from different probes (e.g Refs\;\cite{SNH0, DESYR3, Planck}). While the required precision on $H_0$ is also well within current constraints, it does not span the $H_0$ tension.

It is reassuring that the precision to which we need to know $H_0$ and $\Omega_{\rm m}$ is much larger than the expected constraints from standard sirens on these parameters. But we only have one Universe and one set of observations. As demonstrated in the previous section, the probability for large lensing biases can still be high, even when we are close to being correct to the input cosmology. 

Testing the required simulation resolution is more challenging, as we can not know what the true magnification distribution should be in this case. We use the full-sky simulations of Ref\;\cite{TK}, using the expected weak lensing uncertainty from each $N_{\rm side}$. We also downsample the $N_{\rm side}=4096$ simulations to $N_{\rm side}=2048$ and $N_{\rm side}=1024$. We find a linear scaling between the average bias and ${\rm log}_2(N_{\rm side})$ (with and without the downsampled simulations), which is largely the same, regardless of the true cosmology. This is due to a constant factor difference between these weak lensing fitting functions at $z\geq1$. For every power of 2 increase in $N_{\rm side}$, the average bias drops by $\sim\!0.1\sigma$ where $\sigma$ is the recovered parameter uncertainty. 

To fully investigate the effect of the simulation resolution on the magnification distribution, and relate this to expected magnifications of real sources,  baryonic effects at small scales are required. The AbacusSummit full simulation suite includes simulations with baryons, and weak lensing maps of these are forthcoming \cite{abacus}. Weak lensing maps for full hydro-dynamical simulations with baryons have been calculated in Ref.\;\cite{MTNG}, but for a single resolution of $N_{\rm side}=12288$. In that work, the authors find that baryons reduce the width of the magnification distribution. If this effect could be probed down to smaller scales / greater resolution, this may indicate a convergence of the width of the magnification distribution with resolution, which would provide a best-guess weak lensing uncertainty. 

A similar result could also be obtained using baryonic feedback effects in the halo model used to numerically calculate $\sigma_{\rm WL}(z\,|\,\bm{\theta})$. A feedback model, HMcode2020\_feedback, is already included in \textsc{camb} \cite{hmcode}, but this model was not used in this work due to a divergence in the variance of the power spectrum at small scales.

\subsubsection{Other approaches}

We consider two methods to address imperfect knowledge on the weak lensing uncertainty. The first assumes we are confident in the modelling of $\sigma_{\rm WL}(z)$ and know how it varies with cosmology, but are ignorant of the true values of cosmological parameters. This could be achieved either through improved modelling of small scales in numerical codes such as \textsc{camb}, or a thorough investigation on how the width of the magnification distribution varies across high resolution full-sky simulations using different assumed cosmologies, such as is available with AbacusSummit. The generation of $\sigma_{\rm WL}(z\,|\,\bm{\theta})$ is added as an ingredient to the MCMC analysis with each new sampling of $h$, $\Omega_{\rm m}$ and any other parameters being constrained. This does not guarantee results are unbiased, but removes the sensitivity to an assumed cosmology.

Alternatively, we are not confident in our modelling of $\sigma_{\rm WL}(z)$, and we do not know how it varies with cosmology. Here, we consider introducing an extra parameter into the fitting which gives an amplitude modification to the variance, similar to that given in Ref.\;\cite{LSST_hyperparam}. 
\begin{equation}
    \sigma_{\rm WL}'(z) = A \, \sigma_{\rm WL}(z) \, .
    \label{eq:A}
\end{equation}
A realistic prior on $A$ is needed. The fitting will try to increase the variance as much as it can, as this will minimise $({\rm data}-{\rm model})^2/{\rm variance}$. The prior on $A$ could be based on the difference in predictions for the weak lensing uncertainty across simulations or cosmological parameters. This is a simple choice, and neglects that we are also uncertain on the variation of uncertainty with redshift. A second shape changing parameter could be introduced, but this would also require a well justified prior. This method attempts to find an acceptable bias-variance trade-off, increasing uncertainties within a defined range as you fit the data to a model. It also allows you to jointly constrain cosmology with the form of the weak lensing uncertainty.

If the goal is reliability of parameter constraints, the simplest approach may be to use a large enough weak lensing uncertainty to encompass all possible curves, increasing parameter uncertainties by $\sim\!15\%$ but strongly reducing the probability of obtaining biased results.

\section{Conclusions \label{sec:concl}}

Gravitational waves from binary mergers will be weakly lensed. To address the unknown magnification of each source, a weak lensing uncertainty term must be added to the observations. We aim to investigate the impact of mischaracterising this term on a cosmological analysis.

Our key conclusions are:
\begin{itemize}
    \item Any cosmological analysis using binary mergers at $z\gtrsim 0.5$ not including a weak lensing term has a high probability of being biased by more than $1\sigma$. 
    \item Present weak lensing uncertainty fitting functions could be drastically underestimating the width of the redshift dependent magnification distribution. This is caused by the use of $1\times1\,$deg$^2$ simulations using too small cosmological parameters. We provide a new form for the weak lensing fitting formula in Eq.\;(\ref{eq:sigWL_fit}) using the full-sky ray-tracing simulations of \hyperlink{cite.TK}{T17} in Section \ref{sec:sim}, noting this reduces the probability of a residual lensing bias for $z\leq5.3$ bright standard sirens, such as merging binary neutron stars.
    \item If cosmological parameters are underestimated when generating $\sigma_{\rm WL}(z)$, the probability of parameter constraints biased by more than $1\sigma$ increases by a factor $1.5-3$ compared to the case when these parameters are not underestimated. 
    \item If the magnification distribution is in fact much broader than predicted by small sky-area ray-tracing simulations, but they are used to estimate the weak lensing uncertainty, bias probabilities increase by a factor $3$ ($10$) for BNSs (SMBHBs) compared to the case when the correct $p(\mu)$ is used. The probability of a $1\sigma\,(2\sigma)$ bias in $H_0$ and $\Omega_{\rm m}$ found from BNS mergers is $54\%\,(19\%)$.
    \item     Considering the small absolute value of the bias for BNSs, with a spread of $\Delta b(H_0) \sim 0.25\,{\rm km}\,{\rm s}^{-1}\,{\rm Mpc}^{-1}$, magnification selection effects or incorrect assumptions on the true form of $p(\mu\,|\,z,\bm{\theta})$ are unlikely to impact the efficacy of BNSs on addressing the $H_0$ tension. However should more BSS SMBHBs be detected than anticipated, the absolute bias having a spread $\Delta b(H_0) \sim 5\,{\rm km}\,{\rm s}^{-1}\,{\rm Mpc}^{-1}$ would be significant. An order of magnitude more SMBHBs than considered in these catalogues (from $\sim\!20$ to $\sim\!200$) would be needed for the bias to become a serious consideration.
    \item  Utilising the full potential of bright standard sirens requires accurate modelling of $p(\mu\,|\,z,\bm{\theta})$ using $H_0$,  $\Omega_{\rm m}$ and $\sigma_8$ known to a precision better than $2\,$km$\,$s$^{-1}\,$Mpc$^{-1}$, $0.02$ and $0.07$ respectively. 
    \item The most promising approach to estimate $\sigma_{\rm WL}(z)$ is using high-resolution weak lensing maps, including baryons, over a range of resolutions. A convergence of the width of $p(\mu,z)$ with resolution caused by baryonic feedback effects would provide a reliable estimate for the weak lensing uncertainty. This investigation will be possible in the near-future \cite{abacus}.

\end{itemize}

A similar, independent analysis was performed in Ref.\;\cite{lensbias}. There the true cosmology is that used in the AbacusSummit simulations \cite{abacus} with $h=0.6736$, $\Omega_{\rm m}=0.3138$, similar to the Planck values used in this work, and the weak lensing uncertainty is the fitting function of Ref.\;\cite{LISA_sigWL}. The magnification distributions from AbacusSummit, as seen in Fig.\;1 of Ref.\;\cite{lensbias}, are broader than those from the simulations of \hyperlink{cite.TK11}{T11} from which the WL uncertainty is derived, contributing to their derived bias. Ref.\;\cite{lensbias} include larger lensing magnifications while we limit to $\mu\leq1.75$ to avoid modelling uncertainties. Therefore we could be under-representing the probability of a bias.

With enough data, there is the possibility of constraining the parameters of the weak lensing fitting function in Eq.\;(\ref{eq:sigWL_fit}) jointly with cosmology. We briefly explored this in Eq.\;(\ref{eq:A}), where the amplitude of $\sigma_{\rm WL}(z)$ is treated as a free parameter. This is a hybrid approach, using the information from observations to avoid fixing $\sigma_{\rm WL}(z)$ based only on simulations, and was originally envisioned in Ref.\;\cite{GCAT}. Also, as pointed out in Ref.\;\cite{ULTRA}, the relative abundance of magnified and de-magnified sources reveals the non-Gaussianity of $p(\mu\,|\,z)$. We leave as future work a detailed investigation into the prospects of jointly constraining cosmology and the variance of the magnification distribution.

An interesting consideration is how these results impact delensing prospects. Present delensing studies assume a form for the magnification distribution, which if incorrect could lead to a biased delensing procedure. Future delensing prospects, and the impact of the magnification distribution on delensing, will be explored in a future work. Other future work includes assessing the impact of weak lensing on numerous high redshift dark standard sirens, and investigating the effect of the residual lensing bias for analyses focusing on high redshift SMBHBs to study, for example, modified gravity \cite{SS_MG,SS_MG1,Corman}.

\acknowledgements

We thank Alberto Mangliagli for use of their SMBHB catalogues. CTM thanks Ryuichi Takahashi for helpful discussion and feedback, and Jack Elvin-Poole and Pierre Burger for useful discourse. CTM is supported by a Science and Technology Facilities Council (STFC) Studentship Grant. AT is supported by a STFC Consolidate Grant. M. H. acknowledges the support of the Science and Technology Facilities Council (Grant Ref ST/V005634/1).
The python packages \textsc{numpy}, \textsc{scipy}, \textsc{matplotlib}, \textsc{getdist}, \textsc{emcee}, \textsc{gwfast} and \textsc{tensiometer} have been used in this work. For the purpose of open access, the authors have applied a Creative Commons Attribution (CC BY) licence to any Author Accepted Manuscript version arising from this submission.

\appendix

\section{Lensed distance uncertainty \label{app:sigdL}}

Consider two cases, a source at $d_L$ with no magnification ($\mu=1$), and the same source detected by the same detector but with a magnification $\mu$.

From Fisher formalism,
\begin{align}
    \sigma_{d_L} &= \frac{1}{\sqrt{\Gamma_{d_L}}} \, ,\\
    \Gamma_{d_L} &= \left(\frac{\partial\tilde{h}(f)}{\partial d_L} \mid \frac{\partial\tilde{h}(f)}{\partial d_L}\right) \, ,\\
    \Gamma_{d_L} &= 4\int_{f_{\rm min}}^{f_{\rm max}} \frac{1}{S_{h}(f)}\frac{\partial\tilde{h}^*(f)}{\partial d_L} \frac{\partial\tilde{h}(f)}{\partial d_L} \, df \, .
\end{align}
For simplicity assume
\begin{equation}
    \tilde{h}(f) = \frac{A}{d_L}M_{z}^{5/6}f^{-7/6}e^{i\bm{\Phi}(f)} \, .
\end{equation}
Then
\begin{align}
\Gamma_{d_L} &= \int_{f_{\rm min}}^{f_{\rm max}} \frac{1}{S_{h}(f)}\frac{A^2}{d_L^4}M_{z}^{5/3}f^{-7/6} \, df \, ,\\
\sigma_{d_L} &= \frac{d_L^2}{A}\left(M_{z}^{5/3}\int_{f_{\rm min}}^{f_{\rm max}} \frac{1}{S_{h}(f)}f^{-7/6} \, df\right)^{-1/2} \, .
\end{align}
Hence, the uncertainty on $d_L$
\begin{equation}
    \sigma_{d_L} \propto d_L^2 \, .
\end{equation}
Therefore, using 
\begin{equation}
    d_L' = \frac{d_L}{\sqrt{\mu}} \, ,
\end{equation}
where $'$ corresponds to lensed quantities,
\begin{align}
    \frac{\sigma_{d_L}'}{(d_L')^2} &= \frac{\sigma_{d_L}}{d_L^2} \, , \\
    \sigma_{d_L}' &= \frac{\sigma_{d_L} \left(d_L / \sqrt{\mu}\right)^2}{d_L^2} \, , \\
    \sigma_{d_L}' &= \frac{\sigma_{d_L}}{\mu} \, .
\end{align}
This relation has been verified by passing a catalogue, both before and after lensing, to the GW Fisher parameter uncertainty code GWFish \citep{GWFISH}.

\section{Other ground-based detector networks \label{app:dets}}

\subsection{ET and CE}

If there is only one CE like detector in the network, the sky localisation is degraded. We use a single catalogue of 500 sources to test this case, finding a factor 2 increase in the probability of exceeding $1 \sigma$ when the Planck curve in Fig.\;\ref{fig:sigmu} is used for $\sigma_{\rm WL}(z)$ and the true cosmology is given by the Pantheon+ $\Lambda$CDM constraints. There is a factor 5 increase in bias probabilities when \hyperlink{cite.lens_sel}{CT21} is used for $\sigma_{\rm WL}(z)$ and Planck is the true cosmology. The residual lensing bias is less impactful due to the degraded measurement uncertainties, but still significant.

\subsection{ET and Voyager}

Here we consider a combination of Einstein Telescope and a next generation LVK detector \cite{LIGO_FUTURE}, to aid sky localisation. Due to the EM counterpart selection effects, such as a sky localisation $<20\,$deg$^{2}$ and a distance uncertainty $<30\%$, the redshift of BSS BNS from this detector combination are $\leq0.5$. Selection effects now cause a small bias of $b(h)=-0.0001$ and $b(\Omega_{\rm m})=0.0013$.

If we relax the assumption on the sky localisation to $<200\,$deg$^{2}$, the bias due to selection effects increases by a factor of 5 due to the inclusion of higher redshift sources. But because the uncertainty on $h$ and $\Omega_{\rm m}$ when being jointly constrained by BSS observed in this GW detector network is large, the probability of the residual lensing bias exceeding $1\sigma$ is $0.04$ even when no weak lensing uncertainty term is included. The measurement uncertainty dominates.

We can conclude that the impact of residual lensing biases are most important for high-redshift binaries observed by LISA, or BNS mergers observed through a ground-based network of two or more 3G detectors.

\section{Residual lensing bias in extensions to \texorpdfstring{$\Lambda$CDM}{LCDM} \label{app:cosmo}}

\subsection{Curvature}

\begin{figure}
    \centering
    \includegraphics{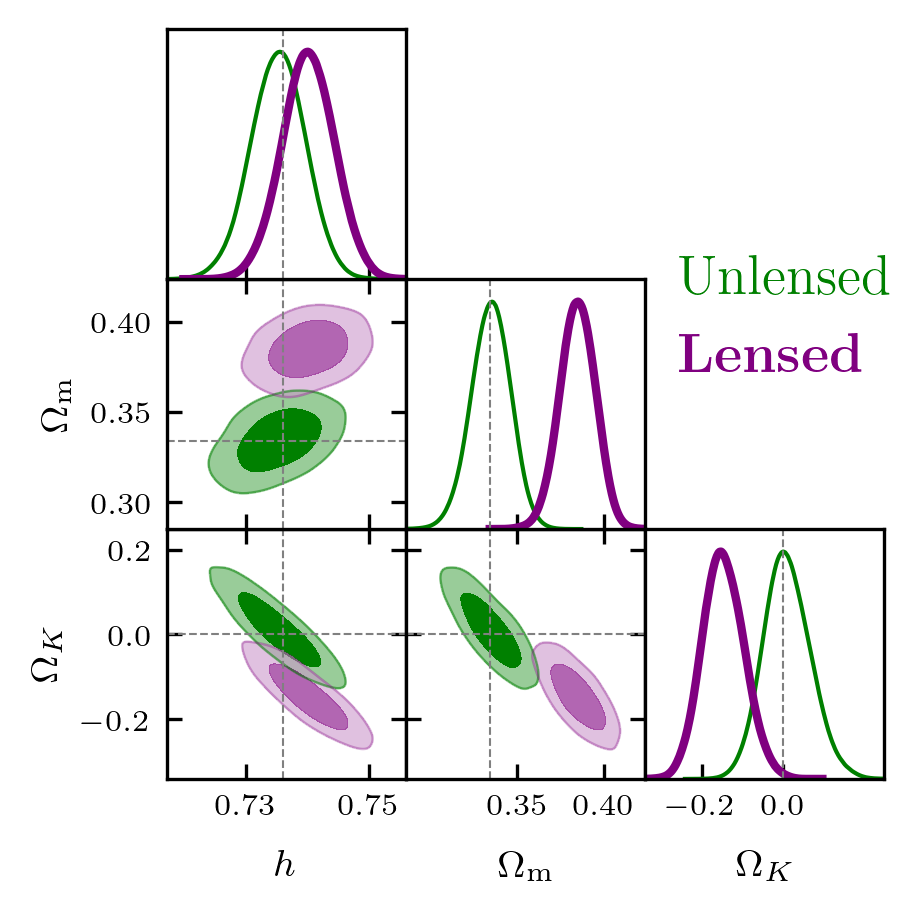}
    \caption{Similar to Fig.\;\ref{fig:lens_bias} but for the $k\Lambda$CDM cosmological model, where we are also constraining curvature. In this case lensing leads to a $2.7\sigma$ bias in this parameter space, and a $>2\sigma$ preference for negative curvature.}
    \label{fig:biask}
\end{figure}

We consider the scenario where the true cosmology is the Pantheon+ $\Lambda$CDM best fit values, and the weak lensing uncertainty is derived using Planck 2018 parameters.  The source population are BNSs observed by the 3G detector network. We find the mean parameter biases caused by magnification selection effects are $\overline{b}(h)=0.0004$, $\overline{b}(\Omega_{\rm m}) = 0.0161$, $\overline{b}(\Omega_k) =-0.03$. We find the probabilities $P(b(h) \geq \pm 0.0025)=0.86$, $P(b(\Omega_{\rm m}) \geq \pm0.015)=0.60$, $P(b(\Omega_{k}) \geq \pm0.1)=0.52$. It can be seen that, for $\Omega_{k}$ particularly, the probability of a residual lensing bias that is both large in terms of the parameter, and large in terms of the parameter uncertainties, is high. The probability of a discordance in the data set $\geq1\sigma\,(\geq2\sigma)$  is $0.19\,(0.02)$. An example of the lensing bias in this cosmological model can be seen in Fig.\;\ref{fig:biask}.

\subsection{Dark Energy}

\begin{figure}
    \centering
    \includegraphics{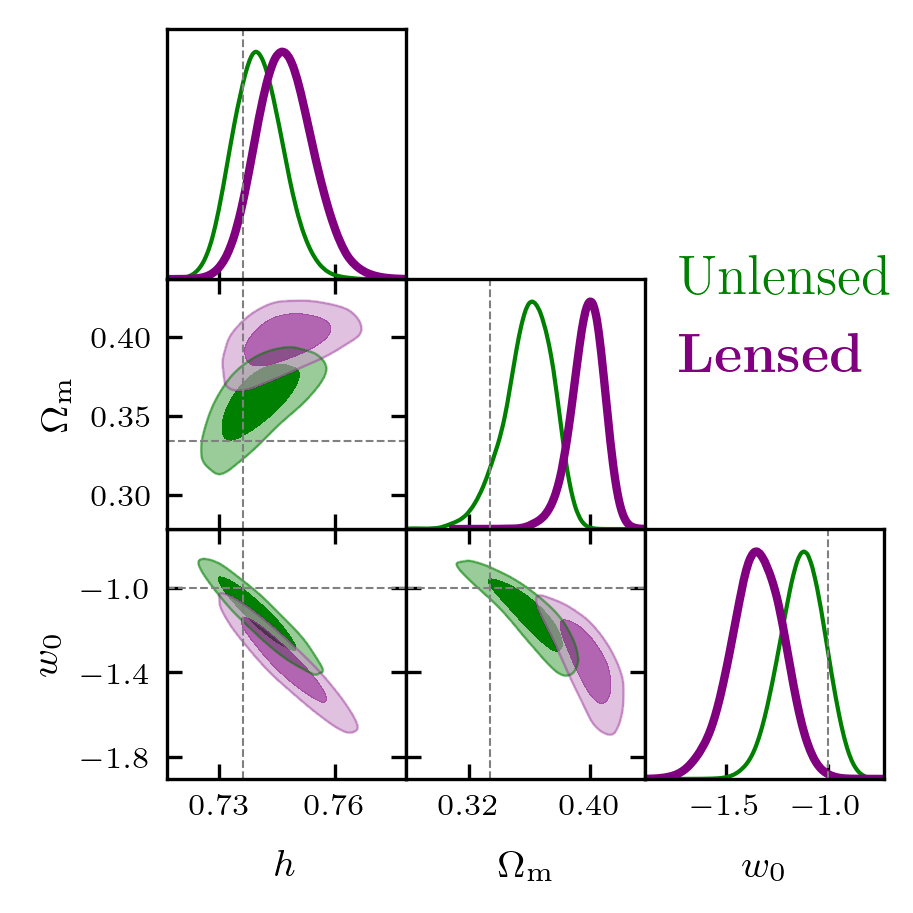}
    \caption{Similar to Fig.\;\ref{fig:lens_bias} but for the $w_0$CDM cosmological model, where we are also constraining a constant dark energy equation of state. In this case lensing leads to a $1.2\sigma$ bias in this parameter space, and a $>2\sigma$ preference for a phantom dark energy.}
    \label{fig:biasw}
\end{figure}

We use the same scenario as in the case of curvature---Pantheon+ $\Lambda$CDM for the true cosmology and Planck to generate the assumed weak lensing uncertainty. We find the mean parameter biases caused by magnification selection effects are $\overline{b}(h)=0.0007$, $\overline{b}(\Omega_{\rm m}) = 0.0236$, $\overline{b}(w_0) =-0.0574$. We find the probabilities $P(b(h) \geq \pm0.0025)=0.72$, $P(b(\Omega_{\rm m}) \geq \pm0.015)=0.77$, $P(b(w_0) \geq \pm0.1)=0.54$. Once again, the parameter extension to $\Lambda$CDM causes significant biases. The probability of a discordance in the data set $\geq1\sigma\,(\geq2\sigma)$  is $0.54\,(0.01)$. An example of the lensing bias in this cosmological model can be seen in Fig.\;\ref{fig:biasw}.


%
\end{document}